\documentclass[10pt, twocolumn]{article}
\usepackage[numbers]{natbib}
\usepackage[a4paper]{geometry}
\usepackage[myheadings]{fullpage}
\usepackage{fancyhdr}
\usepackage{lastpage}
\usepackage{graphicx, wrapfig, subcaption, setspace, booktabs}
\usepackage[T1]{fontenc}
\usepackage[font=small, labelfont=bf]{caption}
\usepackage{fourier}
\usepackage[english]{babel}
\usepackage{sectsty}
\usepackage{url}
\usepackage{lipsum}
\usepackage{tgbonum}
\usepackage{xcolor}
\usepackage{booktabs}
\usepackage{multirow}
\usepackage[normalem]{ulem}
\useunder{\uline}{\ul}{}
\usepackage{tabularx}
\usepackage[utf8]{inputenc}
\usepackage{abstract}
\usepackage{enumitem}
\usepackage{hyperref}
\usepackage{amsfonts}

\usepackage[most]{tcolorbox}
\newtcolorbox{rqframe}[1][]{
  enhanced,
  arc=2pt,
  outer arc=2pt,
  colback=white,
  boxrule=1pt,
  #1
}

\newcommand{\mydate}{November, 2019}

\usepackage{geometry}
\newcommand\dltpy{DLTPy}

\newcommand\todointernal[1]{}
\newenvironment{notsw}[1]{#1}

\newcommand{\HRule}[1]{\rule{\linewidth}{#1}}
\newcommand{\keywords}[1]{\textbf{\textit{Index terms---}} \textit{#1}}

\begin{document}

\fontfamily{cmr}\selectfont
\title{ \normalsize \textsc{}
		\HRule{0.5pt} \\
		\Large \textbf{\uppercase{\dltpy: Deep Learning Type Inference of Python Function Signatures using Natural Language Context}}
		\HRule{2pt}}
		
\author{\small{}
	    Casper Boone \\
	    \small Delft University of Technology\\
	    \texttt{c.c.boone@student.tudelft.nl}
        \and
        \small{}
	    Niels de Bruin \\
	    \small Delft University of Technology\\
	    \texttt{d.j.m.debruin@student.tudelft.nl}
        \and
        \small{}
	    Arjan Langerak \\
	    \small Delft University of Technology\\
	    \texttt{a.c.langerak@student.tudelft.nl}
        \and
        \small{}
	    Fabian Stelmach \\
	    \small Delft University of Technology\\
	    \texttt{f.p.stelmach@student.tudelft.nl}
}

\date{\mydate}

\twocolumn[
\begin{@twocolumnfalse}
\maketitle
\begin{abstract}
Due to the rise of machine learning, Python is an increasingly popular programming language.
Python, however, is dynamically typed.
Dynamic typing has shown to have drawbacks when a project grows, while at the same time it improves developer productivity.
To have the benefits of static typing, combined with high developer productivity, types need to be inferred.
In this paper, we present \dltpy{}: a deep learning type inference solution for the prediction of types in function signatures based on the natural language context (identifier names, comments and return expressions) of a function.
We found that \dltpy{} is effective and has a top-3 F1-score of 91.6\%.
This means that in most of the cases the correct type is within the top-3 predictions.
We conclude that natural language contained in comments and return expressions are beneficial to predicting types more accurately.
\dltpy{} does not significantly outperform or underperform the previous work NL2Type for Javascript, but does show that similar prediction is possible for Python.
\\
\end{abstract}
\end{@twocolumnfalse}
]

\section{Introduction}

Programming languages with dynamic typing, such as Python or JavaScript, are increasingly popular. In fact, supported by the increasing use of machine learning, Python is currently the top programming language in the IEEE Spectrum rankings \cite{2019Interactive:Languages}. Dynamically typed languages do not require manual type annotations and only know the types of variables at run-time. They provide much flexibility and are therefore very suitable for beginners and for fast prototyping. There are, however, drawbacks when a project grows large enough that no single developer knows every single code element of the project. At that point, statically typed languages can check certain naming behaviors based on types automatically whereas dynamic typing requires manual intervention. While there is an ongoing debate on static vs. dynamic typing in the developer community \cite{Ray2017AGitHub}, both excel in certain aspects \cite{Meijer2004StaticLanguages}. There is scientific evidence that static typing provides certain benefits that are useful when software needs to be optimized for efficiency, modularity or safety \cite{Vitousek2014DesignPython}. The benefits include better auto-completion in integrated development environments (IDEs) \cite{Malik2019NL2Type:Information}, more efficient code generation \cite{Vitousek2014DesignPython}, improved maintainability \cite{Hanenberg2014AnMaintainability}, better readability of undocumented source code \cite{Hanenberg2014AnMaintainability}, and preventing certain run-time crashes \cite{Malik2019NL2Type:Information}. It has also been shown that statically typed languages are "less error-prone than functional dynamic languages" \cite{Ray2017AGitHub}.

Weakly typed languages, such as JavaScript, PHP or Python do not provide these benefits. When static typing is needed, there are usually two solutions available. Either using optional type syntax within the language or to use a variant of the language, which is essentially a different language, that does have a type system in place. For JavaScript, there are two often used solutions: Flow \cite{2014Flow}, which uses type annotations within JavaScript, and TypeScript \cite{2012TypeScript}, a JavaScript variant. PHP offers support for type declarations since PHP 7.0 \footnote{https://www.php.net/manual/en/migration70.new-features.php}, and checks these types at run-time. A well-known PHP variant that has strong typing is HackLang \cite{2014HackLang}, by Facebook. Python has support for typing since Python 3.5 \footnote{https://www.python.org/dev/peps/pep-0484/}. It does not do any run-time checking, and therefore these types do not provide any guarantees if no type checker is run. The type checker mypy \cite{2012Mypy} is the most used Python type checker.

Type inference for Python has been addressed from multiple angles \cite{Xu2016PythonSupport,Salib2004FasterStarkiller,MacLachlan1992TheLisp,Hassan2018MaxSMT-Based3c,Maia2012APython}. However, these solutions require some manual annotations to provide accurate results. Having to provide manual annotations is one of the main arguments against static typing because this lowers developer productivity.

In an attempt to mitigate this need for manual annotations and support developers in typing their codebases, we present \dltpy: a deep learning type inference solution based on natural language for the prediction of Python function types. Our work focuses on answering the question of how effective this approach is.

\dltpy{} follows the ideas behind NL2Type \cite{Malik2019NL2Type:Information}, a similar learning-based approach for JavaScript function types. Our solution makes predictions based on comments, on the semantic elements of the function name and argument names, and on the semantic elements of identifiers in the return expressions. The latter is an extension of the ideas proposed in \cite{Malik2019NL2Type:Information}. The idea to use natural language contained in the parameter names for type predictions in Python is not new, Zhaogui Xu et al. already used this idea to develop a probabilistic type inferencer \cite{Xu2016PythonSupport}. Using the natural language of these different elements, we can train a classifier that predicts types. Similar to \cite{Malik2019NL2Type:Information} we use a recurrent neural network (RNN) with a Long Short-Term Memory (LSTM) architecture \cite{Hochreiter1997LongMemory}.

Using 5,996 open source projects mined from GitHub and Libraries.io that are likely to have type annotations, we train the model to predict types of functions without annotations. This works because code has been shown to be repetitive and predictable \cite{Hindle2012OnSoftware}. We make the assumption that comments and identifiers convey the intent of a function \cite{Malik2019NL2Type:Information}.

We train and test multiple variants of \dltpy{} to evaluate the usefulness of certain input elements and the success of different deep learning models. We find that return expressions are improving the accuracy of the model, and that including comments has a positive influence on the results. 

\begin{notsw}
\dltpy{} predicts types with a top-3 precision of 91.4\%, a top-3 recall of 91.9\%, and a top-3 F1-score of 91.6\%. \dltpy{} does not significantly outperform or underperform the previous work NL2Type \cite{Malik2019NL2Type:Information}.
\end{notsw}

This paper's contributions are three-fold:
\begin{enumerate}
    \item A deep learning network type inference system for inferring types of Python functions
    \item Evaluation of the usefulness of natural language encoded in return expressions for type predictions
    \item Evaluation of different deep learning models that indicates their usefulness for this classification task
\end{enumerate}

\begin{figure*}[h]
\centering
\includegraphics[width=\textwidth]{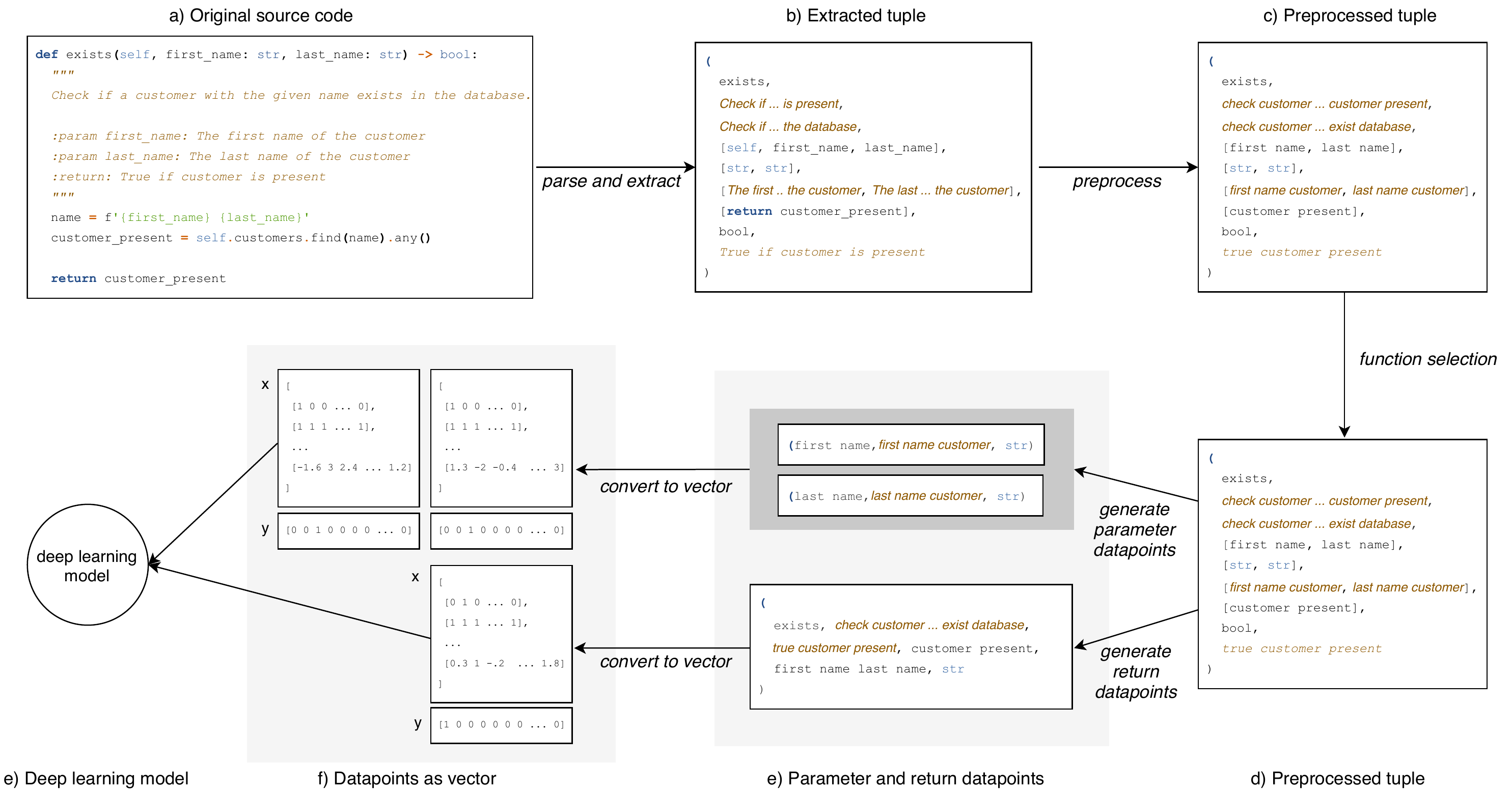}
\caption{Overview of the process of training from annotated source code.}
\label{figure:pipeline}
\end{figure*}

\section{Method} \label{method}
\dltpy{} has two main phases: a training phase and a prediction phase. In this section, we first describe the steps involved in the training process, and then discuss how prediction works, given the trained model. The training process consists of multiple steps. First, we extract relevant training data from Python projects (section \ref{method:extract}). Next, we preprocess the training data by for instance lemmatizing the textual parts of the data (section \ref{method:preprocess}). The preprocessed training data is then filtered and only relevant functions are selected (section \ref{method:selection}). Then, we generate input vectors using word embeddings and one-hot encoding (section \ref{method:vector}). Finally, we train an RNN (section \ref{method:lstm}). After the training process has completed, the trained RNN can be used to make predictions for function types (section \ref{method:prediction}). 

\subsection{Collecting Data from ASTs} \label{method:extract}
For each Python project in our data set, we want to export relevant parts of functions. Every Python file is parsed to an abstract syntax tree (AST). From this AST, we find the functions within or outside a class in the Python file. For each function, we extract the following elements:
\begin{itemize}
    \item $n_f$: The name of the function
    \item $d_f$: The docstring of the function
    \item $c_f$: The comment of the function
    \item $n_p$: A list of the names of the function parameters
    \item $t_p$: A list of the types of the function parameters
    \item $c_p$: A list of the comments describing function parameters
    \item $e_r$: A list of the return expressions of the function
    \item $t_r$: The return type of the function
    \item $c_r$: The comment describing the return value
\end{itemize}

Together, these elements form the tuple $(n_f, d_f, c_f, n_p, t_p, c_p, e_r, t_r, c_r)$. Figure \ref{figure:pipeline}a shows a code sample, this sample is parsed and the information for the tuple is extracted as described in Figure \ref{figure:pipeline}b. This tuple is similar to the input data used in NL2Type \cite{Malik2019NL2Type:Information}, except for $d_f$ and $e_r$.

$d_f$ is the docstring of the Python function. This docstring often contains a few lines of text describing the working of the function, and sometimes also contains information about the parameters or the return value. In some cases, a more structured format is used, such as ReST, Google, or NumPy style. These formats describe parameters and the return value, separately from the function description. In these cases, we can extract this information for $c_f$, $c_p$, and $c_r$. We extract these comments only if the docstring is one of the structured formats mentioned before.

$e_r$ is a list of return expressions of the function. After the preprocessing step (section \ref{method:preprocess}), this contains a list of all the identifiers and keywords used in the return expressions. The intuition is that often variable names are returned and that these names may convey useful information. 

\subsection{Preprocessing} \label{method:preprocess}
The information in the tuple is still raw natural language text. To capture only the relevant parts of the text, we first preprocess the elements in the tuple. The preprocessing pipeline consists of four steps and is based on the preprocessing stage in \cite{Malik2019NL2Type:Information}:

\begin{enumerate}
    \item \textbf{Remove punctuation, line breaks, and digits} We replace all non-alphabetical characters. Line breaks are also removed to create a single piece of text. We replace a full stop that is not at the end of a sentence with a space. We do this to make sure that, for instance, an object field or function access is not treated as a sentence separator (for example \texttt{object.property} becomes \texttt{object property}).
    \item \textbf{Tokenize} We tokenize sentences using spaces as a separator. Before tokenization, the underscores in snake case and camel case identifiers are converted to a space-separated sequence of words. 
    \item \textbf{Lemmatize} We convert all inflected words to their lemma. For example, ``removing'' and ``removed'' become ``remove''.
    \item \textbf{Remove stop words} We remove stopwords (such as ``was'', ``be'', ``and'', ``while'' and ``the'' \footnote{See https://gist.github.com/sebleier/554280 for a full list of stopwords}) from the sentences because these words are often less relevant and thus more importance can be given to non-stopwords. This step is not included in the pipeline for identifiers (function names, parameter names, and return expressions), considering that in the short sentences these identifiers form, stopwords are more relevant.
\end{enumerate}

An example of a preprocessed tuple is shown in Figure \ref{figure:pipeline}c.

\subsection{Function Selection} \label{method:selection}
After collecting and preprocessing the function tuples, we select relevant functions. We filter the set of functions on a few criteria.

First, a function must have at least one type in $t_p$ or it must have $t_r$, otherwise, it cannot serve as training data. A function must also have at least one return expression in $r_e$, since we do not want to predict the type for a function that does not return anything.

Furthermore, for functions where $n_p$ contains the parameter \texttt{self}, we remove this parameter from $n_p$, $t_p$ and $c_p$, since this parameter has a specific role for accessing the instance of the class in which the method is defined in. Therefore, the name of this parameter does not reflect any information about its type and is thus not relevant.

Finally, we do not predict the types \texttt{None} (can be determined statically) and \texttt{Any} (is always correct). Thus, we do not consider a function for predicting a parameter type if the parameter \texttt{Any}, and a return type if the return type is \texttt{Any} or \texttt{None}.

\subsection{Vector Representation} \label{method:vector}
From the selected function tuples, we create a parameter datapoint for each parameter and a return datapoint. We convert these datapoints to a vector. We explain the structure of these vectors in \ref{method:vector:structure}.  All textual elements are converted using word embeddings (see \ref{method:vector:embeddings}), and types using one-hot encoding (see \ref{method:vector:types}).

\subsubsection{Datapoints and Vector Structure} \label{method:vector:structure}

\renewcommand{\arraystretch}{1.6}

\begin{table}[]
\centering
\scriptsize
\begin{tabular}{l l}
\toprule
\textbf{Length} & \textbf{Features}                                                                                                  \\ \midrule
1               & \begin{tabular}[c]{@{}l@{}}Datapoint Type ( \tiny{{[} 1 0 0 ... 0 {]} } )\end{tabular} \\ \hline
1               & Separator                                                                                                          \\ \hline
6               & Name ($n_{p,i}$)                                                                                                              \\ \hline
1               & Separator                                                                                                          \\ \hline
15              & Comment ($c_{p,i}$)                                                                                                            \\ \hline
1               & Separator                                                                                                          \\ \hline
6              & Padding                                                                                                            \\ \hline
1               & Separator                                                                                                          \\ \hline
12              & Padding                                                                                                            \\ \hline
1               & Separator                                                                                                          \\ \hline
10              & Padding                                                                                                            \\ \hline
\end{tabular}
\caption{Vector representation of parameter datapoint.}
\label{table:vector-param}
\end{table}

\begin{table}[]
\scriptsize
\centering
\begin{tabular}{l l}
\toprule
\textbf{Length} & \textbf{Features}                                                                                                  \\ \midrule
1               & \begin{tabular}[c]{@{}l@{}}Datapoint Type ( \tiny{{[} 1 0 0 ... 0 {]} } )\end{tabular} \\ \hline
1               & Separator                                                                                                          \\ \hline
6               & Function Name ($n_f$)                                                                                                              \\ \hline
1               & Separator                                                                                                          \\ \hline
15              & \begin{tabular}[c]{@{}l@{}}Function Comment ($c_f$) \textit{if present, }\\ \textit{otherwise} Docstring ($d_f$)\end{tabular}          \\ \hline
1               & Separator                                                                                                          \\ \hline
6              & Return Comment ($c_r$)                                                                                                           \\ \hline
1               & Separator                                                                                                          \\ \hline
12              & Return Expressions ($e_r$)                                                                                                            \\ \hline
1               & Separator                                                                                                          \\ \hline
10              & Parameter Names ($n_p$)                                                                                                           \\ \hline
\end{tabular}
\caption{Vector representation of return datapoint.}
\label{table:vector-return}
\end{table}

The format of the input vectors is shown in Table \ref{table:vector-param} for parameter datapoints, and in Table \ref{table:vector-return} for return datapoints. All elements of the features have size 14. This results in a 55 $\times$ 14 input vector.

The lengths of the features are based on an analysis of the features in our dataset. The results are shown in Table \ref{table:feature-lengths}. A full analysis is available in our GitHub repository (see section \ref{evaluation:implementation}).

\begin{table*}[]
\centering
\scriptsize
\begin{tabular}{l l l l l l}
\toprule
\multicolumn{1}{c}{\textbf{Feature}}     & \multicolumn{1}{c}{\textbf{\begin{tabular}[c]{@{}c@{}}Average\\ Length\end{tabular}}} & \multicolumn{1}{c}{\textbf{\begin{tabular}[c]{@{}c@{}}Median\\ Length\end{tabular}}} & \multicolumn{1}{c}{\textbf{\begin{tabular}[c]{@{}c@{}}Max\\ Length\end{tabular}}} & \multicolumn{1}{c}{\textbf{\begin{tabular}[c]{@{}c@{}}Chosen\\ Length\end{tabular}}} & \multicolumn{1}{c}{\textbf{\begin{tabular}[c]{@{}c@{}}Fully Covered\\ Data Points\end{tabular}}} \\ \midrule
Function Name*     & 2.28                    & 2                      & 11                  & 6                      & 99,77\%                            \\ \hline
Function Comment** & 6.31                    & 5                      & 482                 & 15                     & 97,98\%                            \\ \hline
Return Comment     & 8.40                    & 5                      & 533                 & 6                      & 65,98\%                            \\ \hline
Return Expressions & 6.01                    & 3                      & 1810                & 12                     & 89,52\%                            \\ \hline
Parameter names    & 2.84                    & 2                      & 72                  & 10                     & 97,67\%                            \\ \hline
Parameter Name*    & 1.45                    & 1                      & 8                   & 6                      & 99,99\%                            \\ \hline
Parameter Comment  & 6.38                    & 4                      & 2491                & 15                     & 93,46\%                            \\ \hline
\end{tabular}
\caption{The chosen vector lengths of the features.}
\label{table:feature-lengths}
\end{table*}

The datapoint type indicates whether the vector represents a parameter or a vector. A separator is a 1-vector of size 14. For parameter datapoints, padding (0-vectors) is used to ensure that the vectors for both datapoints have the same dimensions.

\subsubsection{Learning Embeddings} \label{method:vector:embeddings}
It is important that semantically similar words result in vectors that are close to each other in the n-dimensional vector space, hence we cannot assign random vectors to words. Instead, we train an embeddings model that builds upon Word2Vec \cite{Mikolov2013EfficientSpace}. Since the meaning of certain words within the context of a (specific) programming language are different than the meaning of those words within the English language, we cannot use pre-trained embeddings.

We train embeddings separately for comments and identifiers. Comments are often long (sequences of) sentences, while identifiers can be seen as short sentences. Similarly to \cite{Malik2019NL2Type:Information}, we train two embeddings, because the identifiers ``tend to contain more source code-specific jargon and abbreviations than comments''.

\begin{notsw}
Using the trained model, we convert all textual elements in the datapoints to sequences of vectors.

For the training itself, all words that occur 5 times or less are not considered to prevent overfitting. Since Word2Vec learns the context of a word by considering a certain amount of neighbouring words in a sequence, this amount of set to 5.
The dimension of the word embedding itself is found by counting all the unique words found in the comments and identifiers and taking the 4th root of the result as suggested in \cite{TensorFlowTeam2017IntroducingColumns}. This results in a recommended dimension of 14.
\end{notsw}

\subsubsection{Representing Types}\label{method:vector:types}
The parameter types and return type are not embedded, however, we also encode these elements as vectors. We use a one-hot encoding \cite{Neter1996AppliedModels} that encodes to vectors of length $|T_{frequent}|$, where $T_{frequent}$ is the set of types that most frequently occur within the dataset. We also add the type ``other'' to $T_{frequent}$ to represent all types not present in the set of most frequently occurring types. We only select the most frequent types because there is not enough training data for less frequent types, resulting in a less effective learning process. The resulting vector for a type has all zeros except at the location corresponding to the type, for example, the type \texttt{str} may be encoded as $[0, 1, 0, 0, ..., 0]$.

We limit the set $T_{frequent}$ to the 1000 most frequent types, as this has shown to be an effective number in earlier work \cite{Malik2019NL2Type:Information}. We show the top 10 of the most frequent types in Table \ref{table:most-frequent-types}.

\begin{table}[]
\centering
\scriptsize
\begin{tabular}{l l r}
\toprule
   & \textbf{Type}          & \textbf{Percentage}     \\ \midrule
1  & \texttt{str}            & 28,9\%            \\ \hline
2  & \texttt{int}            & 11,7\%            \\ \hline
3  & \texttt{bool}           & 10,8\%            \\ \hline
4  & \texttt{float}          &  2,9\%            \\ \hline
5  & \texttt{Dict[str, Any]} &  2,4\%            \\ \hline
6  & \texttt{Optional[str]}  &  2,2\%            \\ \hline
7  & \texttt{List[str]}      &  2,1\%            \\ \hline  
8  & \texttt{dict}           &  1,5\%            \\ \hline   
9  & \texttt{Type}           &  1,4\%            \\ \hline   
10 & \texttt{torch.Tensor}   &  1,3\%            \\ \hline  
\end{tabular}
\caption{The top 10 most frequent types in our dataset.}
\label{table:most-frequent-types}
\end{table}

\subsection{Training the RNN} \label{method:lstm}
Given the vector representations described in section (\ref{method:vector}) we want to learn a function that would map the input vectors $x$ of dimensionality k to one of the 1000 types T, hence that would create the mapping $\mathbb{R}^{x*k}->\mathbb{R}^{|T|}$. To learn this mapping, we train a recurrent neural network (RNN). An RNN has feedback connections, giving it memory about previous input and therefore the ability to process (ordered) sequences of text. This makes it a good choice when working with natural language information.

We implement the RNN using LSTM units \cite{Gers1999LearningLSTM}. LSTM units have been successfully applied in NL2Type \cite{Malik2019NL2Type:Information}, where the choice for LSTMs is made based on the use for classification tasks similar to our problem. We describe the full details of the model in \ref{evaluation:experiments:models}.

\subsection{Prediction using the trained RNN} \label{method:prediction}
After training is done, the model can be used to predict the type for new, unseen, functions. The input to the model is similar to the input during the training phase. This means that first a function needs to be collected from an AST (section \ref{method:extract}), then the function elements need to be preprocessed (section \ref{method:preprocess}, and finally, the function must be represented as multiple vectors for the parameter types and for the return type as described in \ref{method:vector}.

The model can now be queried for the input vectors to predict the corresponding types. The network outputs a set of likely types together with the individual probability of the correctness of these types.

\section{Evaluation}
We evaluate the performance of \dltpy{} by creating an implementation, collecting training data, and training the model based on this data. We judge the results using the metrics precision, recall, and F1-score. We also perform experiments using variants of \dltpy{} for comparison.

\subsection{Implementation} \label{evaluation:implementation}
We implement \dltpy{} in Python. We use GitPython \cite{2008GitPython} for cloning libraries, astor \cite{2012Astor} to parse Python code to ASTs, docstring\_parser \cite{2018Docstring_parser} for extracting comment elements, and NLTK \cite{Bird2009NaturalPython} for lemmatization and removing stopwords. We train two word embedding models using the Word2Vec \cite{Mikolov2013EfficientSpace} library by gensim \cite{Rehurek2010SoftwareCorpora}. The data is represented using Pandas dataframes \cite{McKinney2010DataPython} and NumPy vectors \cite{vanderWalt2011TheComputation}. Finally, the prediction models are developed using PyTorch \cite{Paszke2017AutomaticPyTorch}, a machine learning framework.

We make the source for training and evaluating \dltpy{} publicly available at \url{https://github.com/casperboone/dltpy/}.

\subsection{Experimental Setup}
We collect training data from open-source GitHub projects. We first select projects by looking at \texttt{mypy} \cite{2012Mypy} dependents in two ways.
First, we look at the dependents listed on GitHub's dependency graph \footnote{https://github.com/python/mypy/network/dependents}.
Then, we complete this list with the dependents listed by \texttt{Libraries.io} \footnote{https://libraries.io/pypi/mypy/usage}.
The intuition is that Python projects using the type checker \texttt{mypy} are more likely to have types than projects that do not have this dependency. This results in a list of 5,996 projects, of which we can download and process (section \ref{method:extract}) 5,922 projects. 74 projects fail, for instance due to the unavailability of a project on GitHub.

Together, these projects have 555,772 Python files. 5,054 files cannot be parsed.
The files contain 4,977,420 functions, of which only have 585,413 functions have types (at least one parameter type or a return type).

We complement this dataset with projects for which type hints are provided in the Typeshed repository\footnote{https://github.com/python/typeshed}. These are curated type annotations for the Python standard library and for certain Python packages. We manually find and link the source code of 35 packages, including the standard library. Using retype we insert the type annotations into the source code \cite{2017Retype}. These projects have 29,759 functions, of which 246 have types.

We randomly split this set of datapoints into a non-overlapping training set (80\%) and a test set (20\%). We repeat every experiment 3 times.

All experiments are conducted on a Windows 10 machine with an AMD Ryzen 9 3900X processor with 12 cores, 32 GB of memory, and an NVIDIA GeForce GTX 1080 Ti GPU with 11 GB of memory.

\subsection{Metrics} \label{evaluation:metrics}
We evaluate the performance based on the accuracy of predictions.
We measure the accuracy in terms of precision (the fraction of found results that are relevant), recall (the fraction of relevant results found) and F1-score (harmonic mean of precision and recall).
Because in many contexts (for example IDE auto-completion) it is not necessary to restrict to giving a single good prediction, we look at the top-$K$ predictions. 
Specifically, we collect the three metrics for the top-1, top-2, and top-3 predictions.

We define $p$ as the total number of predictions, $p_{valid}$ as the number of predictions for which the prediction is not \texttt{other} (and thus the model cannot make a prediction). We define the three metrics as follows: 

\begin{itemize}
    \item $top\text{-}K\text{ }precision =     
        \dfrac{
            p_{valid\_correct}
        }{
            p_{valid}
        }
        $
        , \\ where $p_{valid\_correct}$ is the number of valid predictions for which the correct prediction is in the top-$K$
    
    \item $top\text{-}K\text{ }recall =     
        \dfrac{
            p_{valid\_correct}
        }{
            p
        }
        $
        ,
    
    \item $top\text{-}K\text{ }F1 = 
        2 \times
        \dfrac{
            top\text{-}K\text{ }precision \times top\text{-}K\text{ }recall
        }{
             top\text{-}K\text{ }precision + top\text{-}K\text{ }recall
        }
        $
        .
\end{itemize}

\subsection{Experiments and Models} \label{evaluation:experiments}
While we mainly evaluate the performance of \dltpy{} as described in the previous section, we also try, evaluate and compare the results when we use different models or different input data. We train three different models and compare these to the model presented in section \ref{method:lstm}. Also, we evaluate the results of selecting different input elements.

\begin{table*}[]
\centering
\scriptsize
\begin{tabular}{l r r r l}
\toprule
\textbf{Dataset}                            & \multicolumn{1}{l}{\textbf{Size}} & \multicolumn{1}{l}{\textbf{\begin{tabular}[l]{@{}l@{}}Parameter\\ Datapoints \end{tabular}}} & \multicolumn{1}{l}{\textbf{\begin{tabular}[l]{@{}l@{}}Return\\ Datapoints \end{tabular}}} & \textbf{Dimensions} \\ \midrule
1 \tiny{Complete}                                    & 84,661                             & 64,690                                             & 19,971                                          & 55 $\times$ 14      \\ \hline
2 \tiny{Optional parameter and return comment}       & 863,936                            & 719,581                                            & 144,355                                         & 55 $\times$ 14      \\ \hline
3 \tiny{Optional docstring}                          & 1,018,787                          & 719,581                                            & 299,206                                         & 55 $\times$ 14      \\ \hline
4 \tiny{Without return expressions}                  & 84,661                             & 64,690                                             & 19,971                                          & 55 $\times$ 14      \\ \hline
5 \tiny{Without return expressions, lower dimension} & 84,661                             & 64,690                                             & 19,971                                          & 42 $\times$ 14      \\ \bottomrule
\end{tabular}
\caption{The sizes of the datasets used to evaluate \dltpy.}
\label{table:datasets}
\end{table*}

\subsubsection{Models} \label{evaluation:experiments:models}
We implement and train three models to evaluate and compare their performance.
\begin{notsw}
\begin{itemize}
   \item \textbf{Model A} In the first model, we made use of two stacked LSTMs. Both LSTMs have a hidden size of 14. The first LSTM will feed its sequence fully into the second, whereas for the second LSTM we only use the output of the last unit and feed it into a fully connected linear layer. Softmax is applied to generate an approximation of the probability for each type. The model has a total of 37,288 parameters.
    
    \item \textbf{Model B} In the second model, we take an approach similar to Model A: we feed the input vector into a Gated Recurrent Unit (GRU) \cite{Cho2014LearningTranslation} and feed the output to a fully connected linear layer converting into the output types. The model has a total of 11,780 parameters.
    
    \item \textbf{Model C} The third model is the model architecture proposed by the authors of \cite{Malik2019NL2Type:Information}. A single bi-directional LSTM with a hidden layer of size 256 is fed into a fully connected layer with output size 1000. Due to the size of this model and the time it took to train, the training consisted of only 25 epochs, compared to 100 epochs for the other three models. The model has a total of 404,456 parameters.
\end{itemize}
\end{notsw}

\subsubsection{Datasets}

We try and evaluate variations in input data to measure the impact of certain elements in the datapoints. To this purpose, we create five datasets. The size of these datasets is listed in Table \ref{table:datasets}.

\begin{enumerate}
    \item \textbf{Complete} This is the dataset as described in section \ref{method}. All datapoints in this dataset are complete. This means that the parameter datapoints have $c_p$, and the return datapoints have $c_f$, $c_r$, and $e_r$. 
    
    \item \textbf{Optional parameter and return comment} In this dataset we make the parameter and return comment optional. The presence of a docstring is still required. This means that parameter datapoints do not have to have $c_p$ (91,01\%), and the return datapoints do not have to have $c_r$ (84,79\%), but still have $c_f$ (which is either the parsed function comment or the docstring) and $e_r$.
    
    \item \textbf{Optional docstring} In this dataset we make the docstring optional. This means that parameter datapoints do not have to have $c_p$ (91,01\%), and the return datapoints do not have to have $c_f$ and $c_r$ (92,66\%), but still have $e_r$ (51,75\%). This can be seen as a dataset without comments since only 20,52\% of the datapoints in this set have comments. This means that the prediction for parameters is purely based on the parameter name, and for return datapoints, the prediction is based on the function name, the return expressions, and the parameter names.
    
    \item \textbf{Without return expressions} To evaluate the usefulness of including return expressions in the model input, we perform the classification task also without return expressions. In this dataset all vectors representing parts of return expressions are 0-vectors.
    
    \item \textbf{Without return expressions, lower dimension} This dataset is similar to the previous one. In this dataset, however, all vectors representing parts of return expressions are removed, resulting in lower-dimensional input vectors.
\end{enumerate}

\section{Results}
In this section we present the results of \dltpy. First, we show and compare the results of our different experiments as described in section \ref{evaluation:experiments} using the metrics described in section \ref{evaluation:metrics}. Then, we compare the results to previous work, specifically to NL2Type \cite{Malik2019NL2Type:Information}.

\begin{table*}
\scriptsize
\centering
\begin{tabular}{llrrrrrrrrr}
\toprule
\multirow{2}{*}{\textbf{Model}}              &
\multirow{2}{*}{\textbf{Dataset}}              & \multicolumn{3}{c}{\textbf{Top-1}}                                & \multicolumn{3}{c}{\textbf{Top-2}}                                & \multicolumn{3}{c}{\textbf{Top-3}}           \\ \cline{3-11} 
                                                & & \textbf{Prec.} & \textbf{Rec.} & \multicolumn{1}{l|}{\textbf{F1}} & \textbf{Prec.} & \textbf{Rec.} & \multicolumn{1}{l|}{\textbf{F1}} & \textbf{Prec.} & \textbf{Rec.} & \textbf{F1} \\ \midrule
A                               & 1 \tiny{Complete}                                    & 60.3                            & 67.7                           & 63.8                    & 71.7                            & 76.7                           & 74.1                    & 76.8                            & 81.2                           & 79.0                         \\
                                & 2 \tiny{Optional parameter and return comment}       & 51.4                            & 62.0                           & 56.2                    & 65.6                            & 69.7                           & 67.6                    & 72.7                            & 75.7                           & 74.1                         \\
                                & 3 \tiny{Optional docstring}                          & 52.9                            & 62.8                           & 57.4                    & 66.1                            & 70.1                           & 68.0                    & 72.9                            & 75.6                           & 74.2                         \\
                                & 4 \tiny{Without return expressions}                  & 60.2                            & 68.6                           & 64.1                    & 72.8                            & 77.5                           & 75.0                    & 78.2                            & 82.0                           & 80.1                         \\
                                & 5 \tiny{Without return expressions, lower dimension} & {\ul 63.9}                      & {\ul 70.3}                     & {\ul 66.9}              & {\ul 74.8}                      & {\ul 78.6}                     & {\ul 76.6}              & {\ul 80.1}                      & {\ul 83.2}                     & {\ul 81.6}                   \\ \hline
B                               & 1 \tiny{Complete}                                    & 41.2                            & 54.7                           & {\ul 47.0}              & 55.2                            & 63.8                           & 59.2                    & {\ul 61.7}                      & 70.0                           & 65.6                         \\
                                & 2 \tiny{Optional parameter and return comment}       & 34.3                            & 51.0                           & 41.0                    & 44.3                            & 57.2                           & 49.9                    & 54.4                            & 63.6                           & 58.6                         \\
                                & 3 \tiny{Optional docstring}                          & 36.6                            & 52.7                           & 43.2                    & 45.7                            & 57.4                           & 50.9                    & 56.5                            & 64.1                           & 60.1                         \\
                                & 4 \tiny{Without return expressions}                  & {\ul 42.2}                      & {\ul 55.9}                     & 48.1                    & {\ul 55.8}                      & {\ul 64.5}                     & {\ul 59.8}              & 62.2                            & {\ul 70.5}                     & {\ul 66.1}                   \\
                                & 5 \tiny{Without return expressions, lower dimension} & 40.4                            & 54.7                           & 46.5                    & 54.7                            & 63.8                           & 58.9                    & 61.1                            & 69.8                           & 65.2                         \\ \hline
C                               & 1 \tiny{Complete}                                    & {\ul \textbf{81.7}}             & {\ul \textbf{83.2}}            & {\ul \textbf{82.4}}     & {\ul \textbf{88.4}}             & {\ul \textbf{89.2}}            & {\ul \textbf{88.8}}     & {\ul \textbf{91.4}}             & {\ul \textbf{91.9}}            & {\ul \textbf{91.6}}          \\
                                & 2 \tiny{Optional parameter and return comment}       & 69.3                            & 70.0                           & 69.6                    & 78.5                            & 78.3                           & 78.4                    & 83.5                            & 83.1                           & 83.3                         \\
                                & 3 \tiny{Optional docstring}                          & 70.7                            & 71.2                           & 71.0                    & 79.8                            & 79.2                           & 79.5                    & 84.5                            & 83.8                           & 84.1                         \\
                                & 4 \tiny{Without return expressions}                  & 79.1                            & 81.3                           & 80.2                    & 86.6                            & 88.0                           & 87.3                    & 90.0                            & 90.9                           & 90.4                         \\
                                & 5 \tiny{Without return expressions, lower dimension} & 81.0                            & 82.8                           & 81.9                    & 88.1                            & 89.0                           & 88.5                    & 91.2                            & 91.6                           & 91.4                         \\ 
\bottomrule
\end{tabular}
\caption{The evaluation results of \dltpy.}
\label{table:results}
\end{table*}

\subsection{Models} \label{results:models}
We presented three different architectures for our prediction model (section \ref{evaluation:experiments}. In Table \ref{table:results} we present the results. The underlined metric scores indicate the best score for each model. The bold metric scores indicate the best overall score. We use these scores to compare the performance of the models.

Model C clearly outperforms models A and B in all metrics. The top-1 F1-score is 82.4\%, and for the top-3 it is 91.6\%. The top-3 recall is 91.9\%, this can be interpreted as the model predicts the correct type within its first three suggestions in 91.9\% of the cases it was asked to make a prediction for. 

Model B performs poorly compared to model A and C. Using a GRU does not have a positive influence on the results. It is interesting to note that where model C performs best on dataset 1, model B does not benefit from the return expressions and performs better without them (dataset 4). 

The performance of model A lies in between the performance of model B and C. Interestingly, this model performs best on the dataset with the lowest dimensions. Dataset 4 and 5 contain the same data, the only difference is that dataset 4 has an additional separator vector and twelve 0-vectors. The same impact can also be seen for model C when comparing dataset 4 and 5. A possible explanation could be that the learning process can converge faster because there is simply less to learn, however, we have not found a provable cause for this.

\subsection{Input Elements Selection}
We presented the elements of the datapoints of \dltpy{} and four variations (section \ref{evaluation:experiments}) that we compare with. In Table \ref{table:results} we present the results. For this comparison, we look at the results of model C, the best performing model.

The results of dataset 2 and 3 are significantly worse than of dataset 1. This shows that the natural language contained in comments positively influences the type classification task. Interestingly, the performance of dataset 3 is slightly less good than of dataset 2, while dataset 2 contains more comments.

Another observation from the results is that the predictions positively influence from including return expressions in the dataset. We see that the performance is less good when return expressions are not included (datasets 4 and 5) than when they are included (dataset 1).

\subsection{Comparison to previous work}

\begin{table}
\scriptsize
\centering
\begin{tabular}{llrrrrrr}
\toprule
\multirow{2}{*}{\textbf{}}              & \multicolumn{3}{c}{\textbf{Top-1}}                                & \multicolumn{3}{c}{\textbf{Top-3}}                                        \\ \cline{2-7} 
                                                & \textbf{Prec.} & \textbf{Rec.} & \multicolumn{1}{l|}{\textbf{F1}} & \textbf{Prec.} & \textbf{Rec.} & \multicolumn{1}{l}{\textbf{F1}} \\ \midrule
                                                
\dltpy{}                   & 81.7          & \textbf{83.2} & \textbf{82.4} &   91.4            & \textbf{91.9}   & 91.6          \\ 
\dltpy{} \tiny{Dataset 4}  & 79.1                            & 81.3                           & 80.2                               & 90.0                            & 90.9                           & 90.4          \\
NL2Type & \textbf{84.1}    & 78.9          & 81.4          &   \textbf{95.5}   & 89.6          & \textbf{92.5}                   \\ 
\bottomrule
\end{tabular}
\caption{A comparison of \dltpy{} to NL2Type \cite{Malik2019NL2Type:Information}.}
\label{table:results-compare}
\end{table}

In this section, we compare our results against NL2Type \cite{Malik2019NL2Type:Information}. \dltpy{} operates in a similar way on similar data as NL2Type. However, NL2Type predicts types for JavaScript files and \dltpy{} predicts types for Python files. Also, \dltpy{} uses a different input representation (different feature lengths and return expressions as an addition) and a different word embedding size (14 instead of 100). This makes it interesting to compare our results against this work.

From the results, we cannot observe that \dltpy{} significantly outperforms or underperforms NL2type (see Table \ref{table:results-compare}). It is, however, interesting to note that without return expressions (dataset 4), \dltpy{} would underperform. Another interesting observation is that the results did not have a negative impact on the significantly smaller word embedding size.

\section{Conclusion}
Our work set out to study the applicability of using natural language to infer types of Python function parameters and return values.
We present \dltpy{} as a deep learning type inference solution based on natural language for the prediction of these types.
It uses information from function names, comments, parameter names, and return expressions. 

The results show that \dltpy{} is effective at predicting types using natural language information. The top-1 F1-score is 82.4\%, and the top-3 F1-score is 91.6\%. This shows that in most of the cases the correct answer is in the top 3 of predictions. The results show that using natural language information from the context of a function and using return expressions have a positive impact on the results of the type prediction task.

We do not significantly outperform or underperform NL2Type. Without our additions to the ideas behind NL2Type, however, \dltpy{} would underperform. This shows that the main idea behind NL2Type, namely using natural language information for predicting types, is generalizable from JavaScript to Python, but additional information, such as return expressions, is needed to get comparable results.

We identify two threats to the validity of our results. The first threat is that there is no separation of functions between the training and test set on project level. Because functions within the same project are more likely to be similar, this might influence the validity of our results. Also, since the best performing model, model C, has 404,456 parameters and the best performing dataset, dataset 1, has just 84,661 datapoints, which increases the risk of overfitting.

\dltpy{} has limitations that can be improved upon in future work. Dataset 1, the complete dataset, is relatively small. A better data retrieval strategy that goes beyond looking at mypy dependents might result in more data points and thus allows for better training resulting in more accurate results. Furthermore, the predictions of \dltpy{} are currently restricted to the 1000 most frequently used types in Python. Open type predictions would improve the practical use of \dltpy{}, given that there is enough training data available for the types that are less frequent.

\bibliographystyle{IEEEtran}
\setlength{\bibsep}{0em}
\renewcommand*{\bibfont}{\small}

\bibliography{references}

\end{document}